 \documentclass[5p,twocolumn]{elsarticle}
\usepackage{amssymb}
\journal{Biological Sciences in Space}

\begin{document}

\begin{frontmatter}

%% Title, authors and addresses

%% use the tnoteref command within \title for footnotes;
%% use the tnotetext command for the associated footnote;
%% use the fnref command within \author or \address for footnotes;
%% use the fntext command for the associated footnote;
%% use the corref command within \author for corresponding author footnotes;
%% use the cortext command for the associated footnote;
%% use the ead command for the email address,
%% and the form \ead[url] for the home page:
%%
%% \title{Title\tnoteref{label1}}
%% \tnotetext[label1]{}
%% \author{Name\corref{cor1}\fnref{label2}}
%% \ead{email address}
%% \ead[url]{home page}
%% \fntext[label2]{}
%% \cortext[cor1]{}
%% \address{Address\fnref{label3}}
%% \fntext[label3]{}

\title{Tanpopo Cosmic Dust Collector: Silica Aerogel Production and\\Bacterial DNA Contamination Analysis}

%% use optional labels to link authors explicitly to addresses:
%% \author[label1,label2]{<author name>}
%% \address[label1]{<address>}
%% \address[label2]{<address>}

\author[First,Second,Third]{Makoto Tabata\corref{cor1}}
\ead{makoto@hepburn.s.chiba-u.ac.jp}
\cortext[cor1]{Corresponding author.} 
\author[Third]{Yuko Kawaguchi}
\author[Third]{Shin-ichi Yokobori}
\author[Second]{Hideyuki Kawai}
\author[Fourth]{Jun-ichi Takahashi}
\author[First]{\\Hajime Yano}
\author[Third]{Akihiko Yamagishi}

\address[First]{Institute of Space and Astronautical Science (ISAS), Japan Aerospace Exploration Agency (JAXA), Sagamihara 252-5210, Japan}
\address[Second]{Department of Physics, Chiba University, Chiba 263-8522, Japan}
\address[Third]{Department of Molecular Biology, Tokyo University of Pharmacy and Life Science, Hachioji 192-0392, Japan}
\address[Fourth]{NTT Microsystem Integration Laboratories, Atsugi 243-0198, Japan}

\begin{abstract}
Hydrophobic silica aerogels with ultra-low densities have been designed and developed as cosmic dust capture media for the Tanpopo mission which is proposed to be carried out on the International Space Station. Glass particles as a simulated cosmic dust with 30 $\mu $m in diameter and 2.4 g/cm$^3$ in density were successfully captured by the novel aerogel at a velocity of 6 km/s. Background levels of contaminated DNA in the ultra-low density aerogel were lower than the detection limit of a polymerase chain reaction assay. These results show that the manufactured aerogel has good performance as a cosmic dust collector and sufficient quality in respect of DNA contamination. The aerogel is feasible for the biological analyses of captured cosmic dust particles in the astrobiological studies.
\end{abstract}

\begin{keyword}
%% keywords here, in the form: keyword \sep keyword
aerogel \sep cosmic dust \sep DNA \sep PCR \sep astrobiology \sep Tanpopo mission
%% MSC codes here, in the form: \MSC code \sep code
%% or \MSC[2008] code \sep code (2000 is the default)
\end{keyword}

\end{frontmatter}

%%
%% Start line numbering here if you want
%%
% \linenumbers

%% main text
\section*{Introduction}
\label{}
The ``Tanpopo'' (Yamagishi \textit{et al}., 2007, 2009) is a mission to conduct several astrobiological studies on the Japanese Experiment Module (JEM) of the International Space Station (ISS). One of primary goals of the Tanpopo mission is to capture dust particles such as micrometeoroids, particles of terrestrial origin and space debris on the low earth orbit (LEO) and to analyze them on ground after safe retrieval. In order to achieve this goal, we are developing a dust particle collector based on silica aerogel.  The aerogel tiles with captured dust particles will be brought back to the ground for chemical and biological analyses.

Terrestrial life may be ejected into outer space by volcanic eruptions, meteorite impacts and other processes (Kring, 2000; Simkin and Siebert, 1994; Griffin, 2004). In fact, microbes have been collected at altitudes from 3 km to 77 km using balloons, aircrafts and rockets since 1936 (Yang \textit{et al}., 2009 as a review). If terrestrial microbes are found in dust particles captured by the Tanpopo mission on the ISS at the LEO ($\sim $400 km altitude), such a result will support the possibility that microbes may escape from the earth to outer space. The dust particles collected by the Tanpopo aerogel collector will be manually retracted from the recovered aerogel tiles for the analysis of microbial DNA by polymerase chain reaction (PCR). PCR is one of the most sensitive methods for detection of DNA. The method can theoretically detect single DNA molecule (Mullis and Faloona, 1987). The PCR amplification of microbial 16S ribosomal RNA (rRNA) gene has been widely used for molecular phylogenetic studies (Weisburg \textit{et al}., 1991). According to hypervelocity impact experiments in ground laboratories, fragmented or molten aerogels adhered around the surface of impacted dust particles during the deceleration process. Therefore, we have to evaluate the effect of adherent aerogels upon the PCR analysis of the dust particle.

Silica aerogel is an optically transparent solid with very low bulk density and has been used for capturing microparticles ($\mu $m) with hypervelocity (km/s) (Burchell \textit{et al}., 2006 as a review). Lower damage of dust particles upon the impact is expected from the extremely low density of aerogels. Aerogel-based dust collectors were already used in past space missions such as ESA EuReCa (Yano and McDonnellm, 1994), the MPAC experiments by JAXA (NASDA) (Kitazawa \textit{et al}., 2000) and the Stardust mission by NASA (Brownlee \textit{et al}., 2006). While aerogel blocks with a density of 0.03 g/cm$^3$ were used in the MPAC experiment, density gradient (0.01--0.05 g/cm$^3$) aerogels were used in the Stardust mission. As for the Tanpopo mission, we plan to utilize aerogel tiles with a double-layered structure. The double-layered aerogel tile is composed of a main surface layer (0.01 g/cm$^3$) to capture dust particles and a base layer (0.03 g/cm$^3$) to support the main surface layer and also to stop dust particles with higher energy. The development of aerogels with ultra-low densities is the key to the Tanpopo mission to obtain micrometeoroids with low damage. For the PCR analysis of captured microbial DNA, aerogel tiles must not be contaminated with bacterial DNA in the course of manufacturing. Aerogels must not interfere with the PCR reaction, either. In this paper, we describe our production method of silica aerogel, and report the results of the hypervelocity impact experiment and the analysis of microbial DNA contaminations in aerogels by PCR. We also discuss the results in terms of the feasibility for the Tanpopo mission.

\section*{Materials and Methods}
\label{}

\subsection*{A Production Method of Silica Aerogel}
\label{}
\begin{figure}[t] 
\centering 
\includegraphics[width=0.50\textwidth,keepaspectratio]{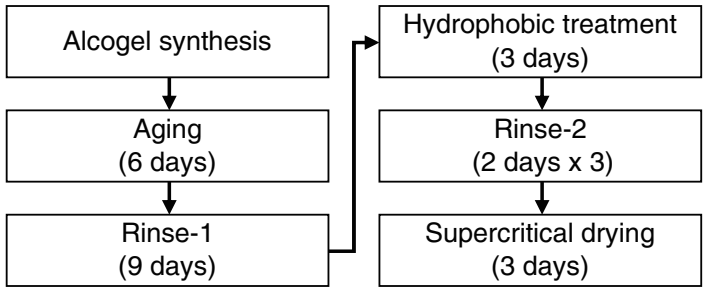}
\caption{A schematic diagram of our production procedure of hydrophobic silica aerogel.}
\label{fig:fig1}
\end{figure}

The schematic diagram of our production procedure of hydrophobic silica aerogel is summarized in Fig. \ref{fig:fig1}, and the detailed procedure of our production method is described below. The production technique is basically same as Adachi \textit{et al}. (1995). We expanded the technique to the density range lower than 0.03 g/cm$^3$ (Tabata \textit{et al}., 2005, 2010). The following production procedure is applied to the production of aerogel of the density range between 0.01 and 0.03 g/cm$^3$.

Aerogels should be cleanly manufactured. Most processes of aerogel production were performed in a clean booth (class 1000). Tools for the aerogel production were soaked in 5\% Extran MA01 (MERCK) for 24 h. They were rinsed with tap water followed by substituting the water layer with ultrapure water. Starting materials are as follows: Methyl Silicate 51 (Fuso Chemical Co., Ltd.), Ethanol (99.5) (special grade chemicals 055-00457, Wako Pure Chemical Industries, Ltd.), 28\% Ammonia Solution (special grade chemicals 016-03146, Wako Pure Chemical Industries, Ltd.), Silazane (Z-6079, Dow Corning Toray Co., Ltd.) and Ethanol (99) (synthetic, non-denatolized, Japan Alcohol Trading Co., Ltd).

\textit{Alcogel synthesis}: Tetramethoxysilane [${\rm Si(OCH}_3{\rm )}_4$] can be easily hydrolyzed, condensed and polymerized in ethanol to produce alcoholic silica gel (alcogel) with the help of a catalyst:
\[ {\rm Si(OCH_3)_4 + 4H_2O \to Si(OH)_4 + 4CH_3OH,} \]
\[ m{\rm Si(OH)_4 \to (SiO_2)}_m {\rm + 2}m{\rm H_2O.} \]
These two successive reactions are the fundamental processes of the alcogel synthesis. Our unique production technique utilizes ``Methyl Silicate 51 (MS51),'' which is commercially available, to simplify the alcogel synthesis process. MS51 is prepared by oligomerizing tetramethoxysilane into oligomers (average degree of polymerization is 4), which has a high content of silica (51\% by weight). At first, two solutions, Solution 1 and 2, were prepared for the alcogel production. Solution 1 was made by adding MS51 to ethanol (99.5). Solution 2 was made by adding ammonia aqueous solution as a base catalyst to ethanol. The Solutions 1 and 2 were mixed in a polyethylene beaker at room temperature and were stirred for 30 sec. The mixed solution was poured into a polystyrene mold and covered with a lid. It took several minutes to complete the gel-formation reaction. We fabricated double-layered alcogels by repeating these operations. After the gelation of the first layer, the mixed solution prepared for the second layer was poured on the first layer. The first and second layers did not mix but were integrally-molded, and were treated as a monolithic tile in the following process. Preparation recipes for aerogels with densities of 0.01 g/cm$^3$ and 0.03 g/cm$^3$ were as follows: MS51:ethanol (99.5):ammonia aqueous solution = 1.43:93.41:5.16 in weight and 4.88:90.16:4.96, respectively.

\textit{Aging}: The synthesized alcogel kept in the polystyrene mold with the lid was placed in an air-tight vessel and aged at room temperature for 6 days to facilitate the formation of a three dimensional network of SiO$_2$. In this aging period, the alcogel shrinks and comes off from the mold. The alcogel became stiff enough to keep its form by itself.

\textit{Rinse-1}: After 6 days, the lid was removed and the air-tight vessel was filled with ethanol (99), so that the alcogel was immersed in ethanol. Then, the vessel was sealed again and put in a water bath and incubated at 35$^\circ$C for 9 days. It is essential to keep the alcogel in ethanol to prevent it from drying. It is important that the alcogel was additionally aged at a constant and good temperature to obtain aerogels with desired densities with high reproducibility.

\textit{Hydrophobic treatment}: Hydroxyl groups on the surface of SiO$_2$ particles are likely to be charged and they can easily react with other ions. They were thus replaced with trimethylsiloxy groups [$-$OSi(CH$_3$)$_3$] by adding hydrophobic reagent, hexamethyldisilazane [((CH$_3$)$_3$Si)$_2$NH]:
\[ {\rm 2(-OH) + ((CH_3)_3Si)_2NH \to 2(-OSi(CH_3)_3) + NH_3.} \]
For the hydrophobic treatment, the alcogel was detached from the mold, transferred into a stainless steel tray with many small holes and immersed into the ethanol used in the rinse-1 process. The hydrophobic reagent was poured into the ethanol. The ratio of the hydrophobic reagent to ethanol was 1:9 in volume. The alcogel was kept in the solution at room temperature for 3 days.

\textit{Rinse-2}: Ammonia generated in the hydrophobic reaction was extracted by ethanol. The alcogel on the tray was transferred into fresh ethanol. The ethanol was replaced by new ethanol every 2 days, 3 times.

\begin{figure}[t] 
\centering 
\includegraphics[width=0.50\textwidth,keepaspectratio]{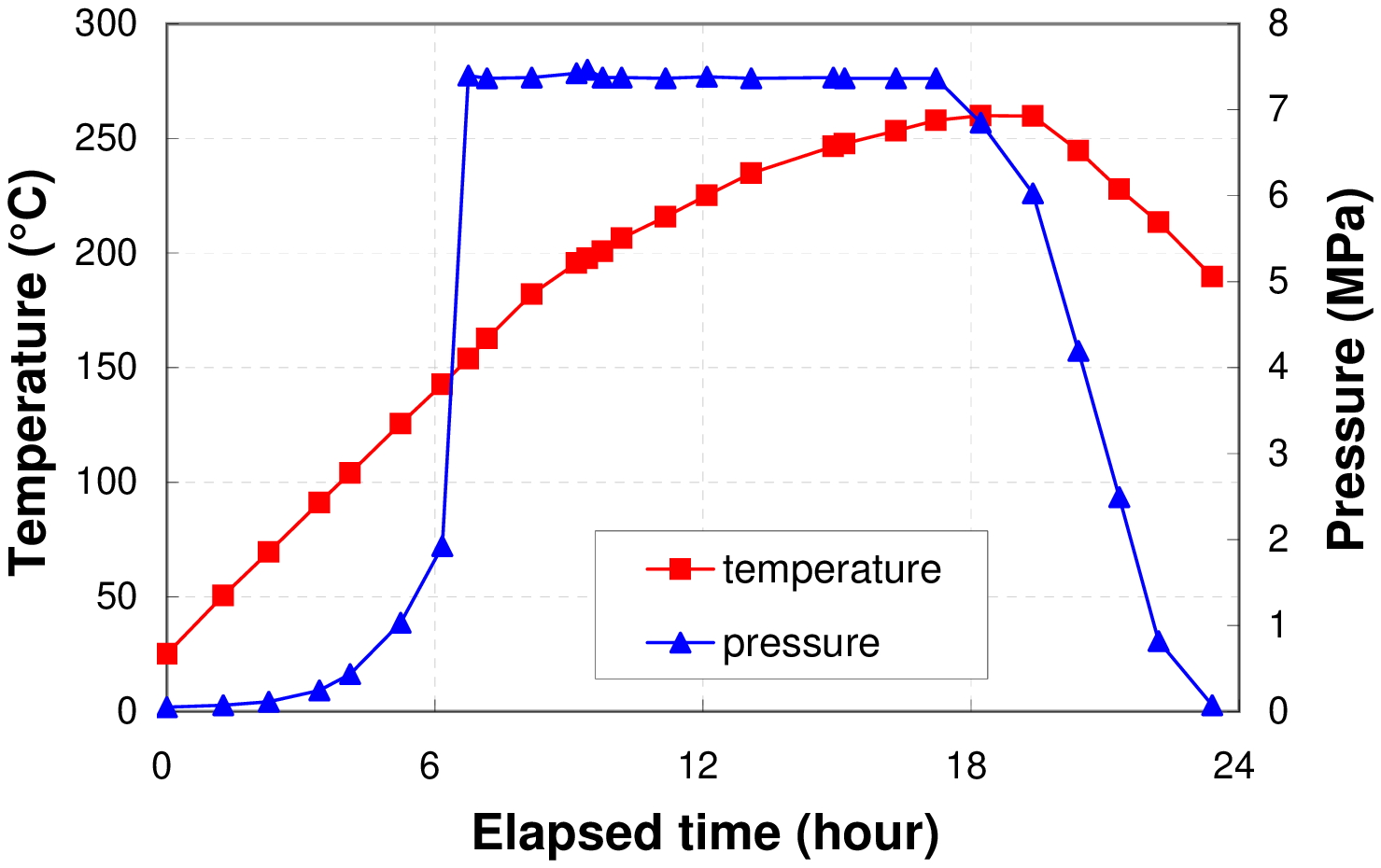}
\caption{Time courses of internal temperature and pressure of the autoclave. The squares and triangles represent the changes in temperature (left axis) and pressure (right axis), respectively.}
\label{fig:fig2}
\end{figure}

\textit{Supercritical drying}: Because the fine structure of silica networks is easily destroyed if the alcogel is dried in air, the alcogel was dried by the supercritical drying method. The critical pressure and temperature of ethanol are 6.4 MPa and 243.1$^\circ$C, respectively. As shown in Fig. \ref{fig:fig2}, pressure and temperature in an ethanol supercritical extraction equipment (semi-automated autoclave) were well controlled around the critical point on the pressure--temperature phase diagram of ethanol. The internal pressure of the sealed and heated autoclave was increased and exceeded the critical pressure (6.4 MPa) by increasing temperature. The pressure was controlled and maintained at 7.3 MPa by opening a pressure control valve. When internal temperature of the further heated autoclave exceeded the critical temperature (243.1$^\circ$C) and reached 260$^\circ$C, pressure reduction was started with the pressure control valve. After the temperature was kept at 260$^\circ$C for 2 h, the heater was turned off. We spent approximately 24 h before the internal pressure returned to the atmospheric pressure, and it took additional 2 days to cool the autoclave to room temperature with injection of nitrogen gas. We spent approximately 1 month to complete all steps for the aerogel production.

\subsection*{Evaluation of Double-layered Silica Aerogel in Hypervelocity Impact Experiment}
\label{}
In order to test the performance of the double-layered aerogel tile as a cosmic dust collector, we have performed a simulated hypervelocity impact experiment using the two-stage light-gas gun at ISAS, JAXA in June 2010. The gun can fire microparticles at a velocity of up to 7 km/s by using a cylindrical case called the ``sabot.'' The drive gas of the launching tube is hydrogen. As a projectile, soda lime glass microspheres (Thermo Fisher Scientific Inc.) with 30.1 $\mu $m in diameter and 2.44 g/cm$^3$ in density were projected at a velocity of 5.97 km/s. A target aerogel (double-layered) was set on an aluminum folder and placed in the vacuum target chamber. The gun sight was taken on the target aerogel by the laser which passed thought the launching tube. The shot was carried out under a vacuum of approximately 10 Pa.

\subsection*{Evaluation of Bacterial Contamination in Silica Aerogel}
\label{}
\textit{Preparation of samples}: The surfaces of an aerogel block were carefully removed aseptically to avoid the possible surface contamination after the production of aerogels. The blocks with various sizes (0.06--0.24 mg) were prepared from the aerogel with a density of 0.009 g/cm$^3$. The DNA sample was prepared according to the method described by Sambrook and Russell (2001). Cells of bacterial species \textit{Deinococcus radiodurans} R1 were cultured in Bacto mTGE medium (Difco) at 30$^\circ$C overnight. The 5 ml of the culture was used for genomic DNA preparation. After centrifugation at 15,000 rpm at 4$^\circ$C for 10 min, the supernatant was discarded and the pellet was suspended into 0.9 ml of T$_{50}$E$_{50}$ buffer (tris (hydroxymethyl) aminomethane 50 mM, EDTA 50 mM, pH 8.0). The 0.1 mL of 10\% (w/v) sodium dodecyl sulfate (SDS) was added to the suspension and they were mixed with a vortex mixer for a few seconds. Samples were frozen in liquid nitrogen for 5 min and then thawed at 56$^\circ$C. This process was repeated five times. The sample was mixed with 40 $\mu $L of 20 mg/mL Proteinase K and incubated at 56$^\circ$C for 2 h. After centrifugation at 15,000 rpm and 4$^\circ$C for 10 min, 750 $\mu $L of the supernatant was recovered. The supernatant was mixed with 750 $\mu $L of isopropanol and 75 $\mu $L of 3 M sodium acetate (pH 7), followed by centrifugation at 15,000 rpm and 4$^\circ$C for 20 min. The pellet was rinsed with 70\% ethanol and dried at room temperature for 20 min.

\textit{PCR conditions}: PCR amplification of approximately 800 bp-long fragments of bacterial 16S rRNA gene was performed with Bacteria-specific primers Eu10F (5'-AGAGTTTGATCCTGGCTCAG-3') and Eu800R (5'-CATCGTTTACGGCGTGGAC-3') (Edwards \textit{et al}., 1989; Hiraishi, 1992). PCR conditions were as follows. The final reaction solution contained 2.5 units LA Taq DNA polymerase HS (TAKARA Bio), 400 $\mu $M each dNTPs, 0.4 $\mu $M each PCR primers, 1 x LA Taq Buffer II (Mg$^{2+}$ free, TAKARA Bio) and 2.5 mM MgCl$_2$. Thermal conditions were 33 cycles of the following thermal cycle consisting of 98$^\circ$C for 10 sec, 45$^\circ$C for 5 sec and 74$^\circ$C for 2 min, after the 94$^\circ$C 1 min treatment as pre-heating. The thermal cycle was followed by 74$^\circ$C for 7 min to complete amplification. The PCR products were checked by 1.5\% agarose gel electrophoresis.

\section*{Results and Discussion}
\label{}

\subsection*{Production of Silica Aerogel}
\label{}
\begin{figure}[t] 
\centering 
\includegraphics[width=0.50\textwidth,keepaspectratio]{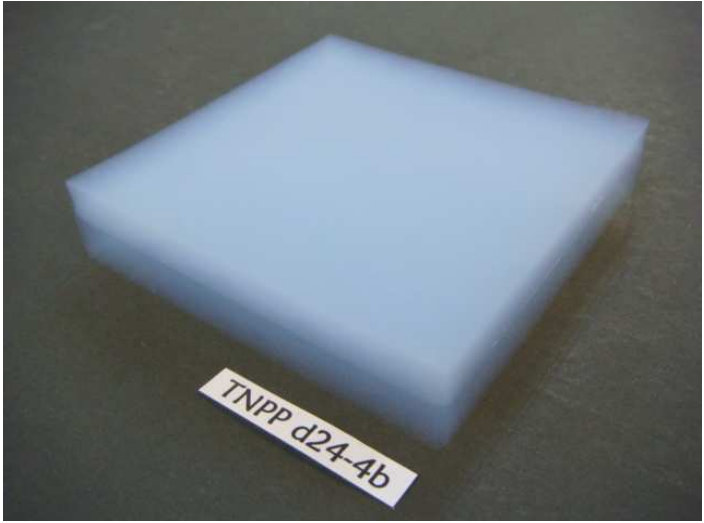}
\caption{A double-layered aerogel sample. The densities of upper and lower layers are 0.01 g/cm$^3$ and 0.03 g/cm$^3$, respectively.}
\label{fig:fig3}
\end{figure}

Fig. \ref{fig:fig3} shows the double-layered aerogel tile fabricated by our method. The tile had a size of approximately 9 cm $\times $ 9 cm $\times $ 2 cm, which will be also utilized as a flight model for the Tanpopo mission. The individuals of the upper layer (1 cm thick, 0.01 g/cm$^3$) and the lower layer (1 cm thick, 0.03 g/cm$^3$) were chemically bound as a monolithic tile. We also produced 0.01 g/cm$^3$ single-layered monolithic tile with the same size however it was generally not resistant to handle. The double-layered aerogel is relatively easy to handle because the lower layer is of adequate strength. The density of our hydrophobic aerogel did not increase due to moisture absorption even if it was kept at a relative humidity of 50\% at 26$^\circ$C for 1 year. At present, the double-layered aerogels are ready for mass production.

\subsection*{Hypervelocity Impact Experiment}
\label{}
The aerogel with an ultra-low density of 0.01 g/cm$^3$ opens up the ideally intact sampling of micrometeoroids. The ultra-low density layer would relax the hypervelocity impact pressure in the process of capturing micrometeoroids. On the other hand, high energy micrometeoroids may entirely penetrate the ultra-low density layer because longer track lengths are expected in the lower aerogel density. The higher density layer at the bottom is expected to stop such micrometeoroids with relatively high energy.

\begin{figure}[t] 
\centering 
\includegraphics[width=0.50\textwidth,keepaspectratio]{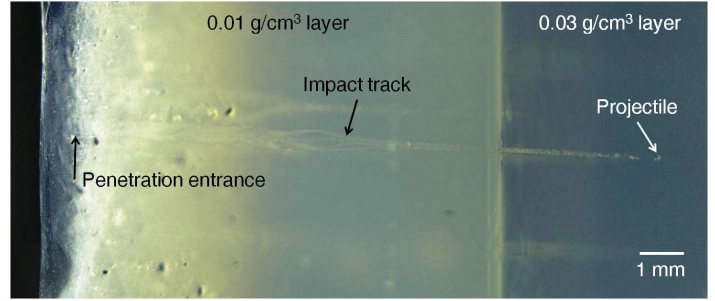}
\caption{The microscope image of the double-layered aerogel used the hypervelocity impact experiment (magnification: $\times $ 30, shot No. 0677, track No. 2). The projectile of soda lime glass with 30 $\mu $m in diameter and 2.44 g/cm$^3$ in density impacted the aerogel from left at a velocity of 5.97 km/s, penetrated the boundary between 2 layers of the aerogel and stopped at the far right of the impact track. The length of the impact track measured 13 mm.}
\label{fig:fig4}
\end{figure}

As a result of the experiment, the projectiles were successfully captured in the double-layered aerogel tile. The aerogel was observed by a digital microscope, VHX-1000 (Keyence Corp.) as shown in Fig. \ref{fig:fig4}. From the impact track observation, the penetration length of the projectile was determined to be 9.4 mm and 3.6 mm in the 0.01 g/cm$^3$ layer and 0.03 g/cm$^3$ layer, respectively. There was little volume change of the projectile. The result suggests that it is possible to capture dust particles with a diameter of 30 $\mu $m using the aerogel with our configuration on the LEO. Thus the novel aerogel as a hypervelocity cosmic dust collection medium is feasible for Tanpopo mission.

\begin{figure}[t] 
\centering 
\includegraphics[width=0.50\textwidth,keepaspectratio]{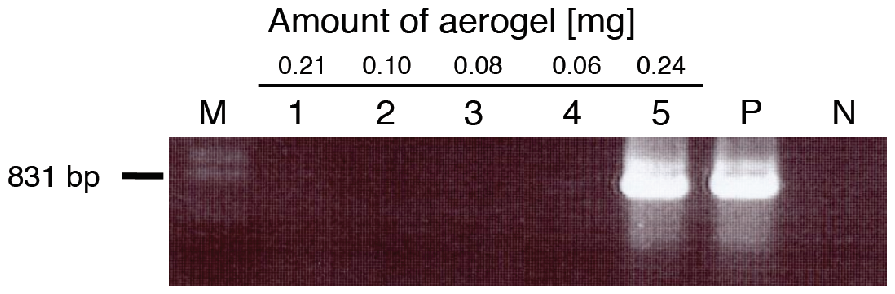}
\caption{Agarose gel electrophoresis of the amplified PCR product of bacterial 16S rRNA gene for evaluation bacterial contamination in the aerogel. The number of each lane indicates weight of aerogels [mg] in each PCR reaction solution. Lanes 5 and P show the PCR amplification with \textit{D. radiodurans} R1 genomic DNA (4.6$\times $10$^{-4}$ $\mu $g for each solution). Lane N (negative control) shows the result of PCR with no genomic DNA. Lane M is for $\lambda $/\textit{Hin}d III + \textit{Eco}R I markers.}
\label{fig:fig5}
\end{figure}

\subsection*{Contamination Check of Microbe DNA}
\label{}
The 16S rRNA is an essential gene for bacteria thus possessed by all of the bacteria.  The 16S rRNA has been routinely analyzed for the environmental bacterial analysis (Tringer \textit{et al}., 2005). The 16S rRNA gene was amplified from the aerogel blocks with different sizes. The PCR product was analyzed with agarose gel electrophoresis. The result of the agarose gel electrophoresis is shown in Fig. \ref{fig:fig5}. No DNA fragment band of the partial bacterial 16S rRNA gene was observed in PCR products with various amounts of the aerogel up to 0.21 mg i.e. 0.02 cm$^3$ as shown in lanes from 1 to 4 in Fig. \ref{fig:fig5}. The result suggests that detectable amount of contaminated DNA does not exist in 0.21 mg of the aerogel sample. Possible inhibitory effect of the PCR reaction caused by the co-presence of the aerogel was also tested. However, the DNA fragment band with the same intensity was observed after the PCR amplification of genomic DNA (4.6 $\times $ 10$^{-4}$ $\mu $g) irrespective of the presence and absence of the aerogel (0.24 mg) (lanes 5 and P of Fig. \ref{fig:fig5}). Some minerals including silica are inhibitor of nucleic acid amplification by PCR (Wilson, 1997). But the aerogel that we have tested did not inhibit the PCR amplification of the gene. The amount of DNA in the aerogel was lower than the detection limit of our PCR analysis.

\subsection*{Concluding Remarks}
\label{}
Hydrophobic silica aerogel with an ultra-low density of 0.01 g/cm$^3$ has been cleanly manufactured, and the double-layered aerogel tile has been successfully developed to test the feasibility for the Tanpopo mission. As a result of the hypervelocity impact experiment, we succeeded in capturing 30 $\mu $m glass projectiles at a velocity of 6 km/s by the double-layered aerogel. The evaluation of bacterial contamination in the aerogel was conducted by PCR. As the result of the investigation of the inhibitory effect, we confirmed that the aerogel does not inhibit PCR. The amount of DNA was lower than the detection limit of PCR analysis. These results support the feasibility of the current aerogel as a medium for capturing micro dust particles in Tanpopo mission on the LEO.

\section*{Acknowledgments}
\label{}
We would like to thank Prof. I. Adachi of High Energy Accelerator Research Organization (KEK) for his assistance in the aerogel production. The hypervelocity impact experiment was supported by the Space Plasma Laboratory, ISAS, JAXA, and we would like to thank Dr. S. Hasegawa for his assistance in the experiment. We are also grateful to our Tanpopo colleagues for fruitful discussions. This work was partially supported by Grant-in-Aid for JSPS Fellows (No. 07J02691 for M.T.) from Japan Society for the Promotion of Science (JSPS).

%% The Appendices part is started with the command \appendix;
%% appendix sections are then done as normal sections
%% \appendix

%% \section{}
%% \label{}

%% References
%%
%% Following citation commands can be used in the body text:
%% Usage of \cite is as follows:
%%   \cite{key}          ==>>  [#]
%%   \cite[chap. 2]{key} ==>>  [#, chap. 2]
%%   \citet{key}         ==>>  Author [#]

%% References with bibTeX database:

%%% \bibliographystyle{model1-num-names}
%%% \bibliography{<your-bib-database>}

%% Authors are advised to submit their bibtex database files. They are
%% requested to list a bibtex style file in the manuscript if they do
%% not want to use model1-num-names.bst.

%% References without bibTeX database:

\end{document}